\documentclass[pra,twocolumn,aps,amssymb,showpacs,superscriptaddress]{revtex4-1}

\usepackage{epsfig}
\usepackage{graphicx}
\usepackage{amsmath} 
\usepackage{amssymb}
\usepackage{enumerate}
\usepackage{hyperref}
\usepackage[normalem]{ulem}
\usepackage{cancel}

\usepackage{color}
\definecolor{rosso}{rgb}{1,0,0}
\definecolor{verde}{rgb}{0,1,0}
\definecolor{blue}{rgb}{0,0,1}
\definecolor{verdescuro}{rgb}{0,0.5,0.5}
\definecolor{rossoscuro}{rgb}{0.7,0.3,0}
\definecolor{bluscuro}{rgb}{0.3,0,0.7}
\definecolor{magenta}{rgb}{1,0,1}

\begin{document}

\title{Josephson current flowing through a nontrivial geometry: \\ The role of pairing fluctuations across the BCS-BEC crossover}

\author{V. Piselli}
\affiliation{School of Science and Technology, Physics Division, Universit\`{a} di Camerino, 62032 Camerino (MC), Italy}
\affiliation{CNR-INO, Istituto Nazionale di Ottica, Sede di Firenze, 50125 (FI), Italy}
\author{L. Pisani}
\affiliation{School of Science and Technology, Physics Division, Universit\`{a} di Camerino, 62032 Camerino (MC), Italy}
\author{G. Calvanese Strinati}
\email{giancarlo.strinati@unicam.it}
\affiliation{School of Science and Technology, Physics Division, Universit\`{a} di Camerino, 62032 Camerino (MC), Italy}
\affiliation{CNR-INO, Istituto Nazionale di Ottica, Sede di Firenze, 50125 (FI), Italy}


\begin{abstract}
A realistic description of the Josephson effect at finite temperature with ultra-cold Fermi gases embedded in nontrivial geometrical constraints (typically, a trap plus a barrier) requires appropriate consideration of pairing fluctuations that arise  in inhomogeneous environments.
Here, we apply the theoretical approach developed in the companion article [Pisani \emph{et al.}, Phys. Rev. B {\bf 108}, 214503 (2023)], where the inclusion of pairing fluctuations 
beyond mean field across the BCS-BEC crossover at finite temperature is combined with a detailed description of the gap parameter in a nontrivial geometry.
In this way, we are able to account for the experimental results on the Josephson critical current,
reported both at low temperature for various couplings across the BCS-BEC crossover and as a function of temperature at unitarity.
Besides validating the theoretical approach of the companion article, our numerical results reveal generic features of the Josephson effect which may not readily emerge from an  
analysis of corresponding experiments with condensed-matter samples owing to the unique intrinsic flexibility of experiments with ultra-cold gases.
\end{abstract}

\maketitle

\section{Introduction} 
\label{sec:introduction}

Most practical applications of superconductors (or, generically speaking, of fermionic superfluids) require a detailed consideration of inhomogeneous environments.
In this context, the theoretical description based on the Ginzburg-Landau (GL) equation for the complex order parameter has proved quite useful \cite{Tinkham-1975}.
However, the GL equation is valid only close enough to the critical temperature $T_{c}$ and for weak inter-particle coupling when the Cooper pair size is much larger than the inter-particle distance.
In fermionic superfluids with strong enough coupling (like ultra-cold Fermi gases) these restrictions are in general violated and the GL equation cannot be applied.
In these cases, one may revert to solving the Bogoliubov-deGennes (BdG) equations \cite{DeGennes-1966}, which are equivalent to the inhomogeneous version of the BCS theory 
developed by Gor'kov \cite{Gorkov-1958} and can, in principle, be applied for any coupling across the BCS-BEC crossover, both at zero \cite{Spuntarelli-2010} and finite \cite{Simonucci-2013} temperature.

In practice, the problem with the BdG equations is twofold.
When solving numerically  these two-component Schr\"{o}dinger-like equations, difficulties arise  in storing the large number of details contained in the single-particle wave-functions (from which the order parameter is eventually obtained through an averaging procedure that washes out most of these details).
In addition, the BdG equations do not take into account  pairing fluctuations, whose consideration is required away from the weak-coupling (BCS) limit
of the BCS-BEC crossover \cite{Phys-Rep-2018}.

In this respect, a first attempt to include exchange and correlation effects in the BdG equations for spatially inhomogeneous superconductors was proposed in Ref.~\cite{Kohn-1988} in terms of Kohn-Sham-type equations,
where the practical challenge is  to include pairing correlations in the exchange-correlation free-energy functional.
It is for this reason that most implementations of this proposal relied on rather pragmatical semi-phenomenological approaches to that  functional \cite{Russmann-2022}.
By a related token, a superfluid local density approximation (SLDA) variant has expressed the energy density functional  at low temperature 
in terms of three phenomenological parameters, that could be determined only at unitarity \cite{Bulgag-2007} and in the weak-coupling (BCS) limit \cite{Magierski-2022} by exploiting known independent  results.
This method has recently been utilized for studying dissipation effects in the context of the Josephson effect \cite{Magierski-2023}.

This article addresses specifically the question of including pairing fluctuations in the BdG equations when the coupling spans the BCS-BEC crossover and the temperature is below 
the superfluid critical temperature $T_{c}$.
This enable us to provide a detailed theoretical account of the experimental results reported in Refs.~\cite{Kwon-2020,DelPace-2021} for the Josephson effect in ultra-cold superfluid Fermi gases 

To this end, we adopt the theoretical approach developed  in Ref.~\cite{PPS-companion-PRB-article}, which is alternative to the approaches  of Refs.~\cite{Kohn-1988}-\cite{Magierski-2023}
and enables us to conveniently deal with spatially inhomogeous fermionic superfluids at \emph{any\/} coupling across the BCS-BEC crossover and at \emph{any\/} temperature below $T_{c}$.
Moreover,  being  based on the many-body Green's functions theory, this novel approach is amenable to further improvements through a ``modular'' inclusion of additional diagrammatic contributions, like the extended Gorkov-Melik-Barkhudarov (GMB) approach in the superfluid phase introduced in Ref.~\cite{Pisani-2018}.

\section{Theoretical approach and methods} 
\label{sec:theoretical_approach}

The present approach includes pairing fluctuations on top of a simplified version of the BdG equations, thereby overcoming the difficulties described in the Introduction.
Specifically, the present approach succeeds in merging: 
(i)  The coarse-graining procedure on the BdG equations introduced  in Ref.~\cite{Simonucci-2014}, which results into the local phase density approximation 
(LPDA) differential  equation for the complex gap parameter $\Delta(\mathbf{r})$;
(ii) A local version of the $t$-matrix approximation  for fermionic superfluids implemented in Ref.~\cite{Pieri-2004}.
The merging of (i) and (ii) eventually transforms the LPDA approach of Ref.~\cite{Simonucci-2014} into the mLPDA approach of Ref.~\cite{PPS-companion-PRB-article},
and yields expressions for the local density and current that take into account beyond-mean-field pairing fluctuations in the presence of spatial inhomogeneities.
These expressions have the property to evolve  with continuity from a fermionic to a bosonic \emph{two-fluid model\/} when the coupling spans the BCS-BEC crossover.
Following the procedures utilized in  Ref.~\cite{Piselli-2020} for the LPDA equation, the ensuing mLPDA equation will explicitly be solved for the Josephson effect in the presence of 
 nontrivial geometrical constraints, like those utilized  experimentally in Refs.~\cite{Kwon-2020,DelPace-2021}.

The LPDA equation for $\Delta(\mathbf{r})$ reads \cite{Simonucci-2014}:
\begin{equation}
\hspace{-0.25cm} \left[ \frac{m}{4 \pi a_{F}} + \mathcal{I}_{0}(\mathbf{r}) +  
\mathcal{I}_{1}(\mathbf{r}) \! \left( \frac{\nabla^{2}}{4m} - i \, \frac{\mathbf{A}(\mathbf{r})}{m} \cdot \nabla \! \right) \! \right]\! \Delta(\mathbf{r}) = 0                                                
\label{LPDA-differential-equation}
\end{equation}
where $\hbar = 1$ and the expressions of the (highly nonlinear) coefficients $\mathcal{I}_{0}(\mathbf{r})$ and $\mathcal{I}_{1}(\mathbf{r})$ are reported in Appendix~\ref{sec:Appendix-A}.
In the context of the BCS-BEC crossover, the scattering length $a_{F}$ of the two-fermion problem that enters Eq.~(\ref{LPDA-differential-equation}) is combined with the Fermi wave vector $k_{F} = (3 \pi^{2} n)^{1/3}$ 
with density $n$, to obtain the dimensionless coupling $(k_{F} a_{F})^{-1}$.
This coupling ranges from $(k_{F}\, a_{F})^{-1} \lesssim -1$ in the weak-coupling (BCS) regime when $a_{F} < 0$, to $(k_{F}\, a_{F})^{-1} \gtrsim +1$ in the strong-coupling (BEC) regime when $a_{F} > 0$, 
across the unitary limit $(k_{F}\, a_{F})^{-1} = 0$ when $|a_{F}|$ diverges \cite{Phys-Rep-2018}.
In this context, the LPDA equation recovers both the GL equation in the BCS limit close to $T_{c}$ and the Gross-Pitaevskii (GP) equation in the BEC limit at low temperature \cite{Simonucci-2014}.
In addition, in the presence of a steady supercurrent, in Eq.~(\ref{LPDA-differential-equation}) one identifies $\mathbf{A}(\mathbf{r}) \rightarrow - \mathbf{Q}_{0}$ where the wave vector $\mathbf{Q}_{0}$ 
contributes the phase $2 \mathbf{Q}_{0} \cdot \mathbf{r}$ to the gap parameter $\Delta(\mathbf{r})$ in a homogeneous environment \cite{DeGennes-1966}.

When solving numerically the LPDA differential equation in the presence of a Josephson barrier, in Ref.~\cite{Piselli-2020} the imaginary part of the LPDA equation (\ref{LPDA-differential-equation}) 
was conveniently replaced by the constraint for the supercurrent to be everywhere uniform.
To satisfy this constraint, an additional local phase $2 \phi(\mathbf{r})$ adds to $2 \mathbf{Q}_{0} \cdot \mathbf{r}$, whose spatial profile acts to compensate for the local variation of the magnitude $|\Delta(\mathbf{r})|$ 
close to the barrier.
Correspondingly, the expressions for the local density and current read \cite{PPS-companion-PRB-article}:
\begin{eqnarray}
n(\mathbf{r}) & = & \frac{2}{\beta} \sum_{n} e^{i \omega_{n} \eta} \! \int \!\!\! \frac{d \mathbf{k}}{(2 \pi)^{3}} \,\, \mathcal{G}_{11}(\mathbf{k},\omega_{n};\mathbf{q}|\mathbf{r})
\label{local-density} \\
\mathbf{j}(\mathbf{r}) & = & \frac{1}{m} \left( \mathbf{Q_{0}} + \nabla \phi(\mathbf{r}) \right) n(\mathbf{r}) 
\nonumber \\
& + & \frac{2}{\beta} \sum_{n} e^{i \omega_{n} \eta} \! \int \!\!\! \frac{d \mathbf{k}}{(2 \pi)^{3}} \frac{\mathbf{k}}{m} \, \mathcal{G}_{11}(\mathbf{k},\omega_{n};\mathbf{q}|\mathbf{r}) \, .
\label{local-current}
\end{eqnarray}
Here,
$\beta = (k_{B} T)^{-1}$ is the inverse temperature ($k_{B}$ being the Boltzmann constant),
$\eta$ a positive infinitesimal,
$m$ the fermion mass,
$\omega_{n} = (2n+1)\pi/\beta$ ($n$ integer) a fermionic Matsubara frequency \cite{FW-1971},
and $\mathbf{k}$ a three-dimensional wave vector.
In addition, $\mathcal{G}_{11}(\mathbf{k},\omega_{n};\mathbf{q}|\mathbf{r})$ is the diagonal (``normal'') single-particle Green's function in the superfluid phase \cite{Schrieffer-1964}, 
which has now to be consistently obtained in the presence of the supercurrent  such that $\mathbf{q} \rightarrow  \mathbf{Q_{0}} + \nabla \phi(\mathbf{r})$ in Eqs.~(\ref{local-density}) and (\ref{local-current})
(where the implicit dependence on $\mathbf{r}$ originates from the barrier) \cite{footnote-phase}.

In the LPDA approach of Ref.~\cite{Piselli-2020}, the Green's function $\mathcal{G}_{11}$ was taken at the mean-field level.
As a consequence, the results obtained therein cannot confidently be extended to the BEC side of the crossover, especially at finite temperature.
Here, we go beyond mean field and include pairing-fluctuation corrections in the expression of $\mathcal{G}_{11}$, in order to span the whole BCS-BEC crossover successfully.
To this end, we resort to the $t$-matrix approximation for the self-energy in the superfluid phase, in the form introduced in Ref.~\cite{Pieri-2004} but now modified so as to account for the presence of a supercurrent.
Once this new expression of $\mathcal{G}_{11}$ (obtained in Ref.~\cite{PPS-companion-PRB-article} and also reported in Appendix~\ref{sec:Appendix-A}) is utilized in Eqs.~(\ref{local-density}) 
and (\ref{local-current}), the LPDA approach of Ref.~\cite{Simonucci-2014} evolves into the (modified) mLPDA approach as proposed in Ref.~\cite{PPS-companion-PRB-article}.
At the same time, the beyond-mean-field value of the chemical potential correctly ranges from the Fermi energy $E_{F} = k_{F}^{2}/(2m)$ in the BCS limit to (half) the binding energy of the dimers 
that form in the BEC limit \cite{Phys-Rep-2018}.

The $t$-matrix approximation was originally considered by Galitskii \cite{Galitskii-1958} for a repulsive dilute Fermi gas with $k_{F} a_{F} \ll 1$, by summing a whole series of ladder diagrams to replace 
the strength of the (contact) inter-particle interaction by the scattering length $a_{F} > 0$. 
Soon after, Gork'ov and Melik-Barkhudarov (GMB) applied this treatment to the case of an attractive inter-particle interaction for which $a_{F} < 0$, 
albeit still in the BCS limit where $k_{F} |a_{F}| \ll 1$ \cite{GMB-1961}.
More recently, Nozi\`eres and Schmitt-Rink extended to the whole BCS-BEC crossover the range of validity of this (non-self-consistent) $t$-matrix approximation for an attractive Fermi gas, 
although only in the normal phase above $T_{c}$ \cite{NSR-1985}.
Several degrees of self-consistency for the $t$-matrix in the normal phase have since been considered and compared with each other \cite{Pini-2019}.
Extensions of the $t$-matrix in the superfluid phase below $T_{c}$ have also been implemented, in both partially self-consistent \cite{Pieri-2004} and fully self-consistent \cite{Haussmann-2007} versions.
Here, we adopt the $t$-matrix approach of Ref.~\cite{Pieri-2004}, also because it fits well with the beyond-$t$-matrix project, that was set up for a homogeneous superfluid in Ref.~\cite{Pisani-2018}
and preliminary extended to the presence of inhomogeneous environments in Ref.~\cite{PPS-companion-PRB-article}.

The LPDA differential equation (\ref{LPDA-differential-equation}) can be solved with reasonable numerical efforts even in the presence of nontrivial geometrical constraints in which the fermionic superfluid is embedded.
Here, we describe the experimental geometry set up of Refs.~\cite{Kwon-2020,DelPace-2021}, which is utilized to obtain the numerical results presented in Sec.~\ref{sec:comparison_experim}.

\section{Geometrical constraints} 
\label{sec:geom_constraints} 

\begin{figure}[t]
\begin{center}
\includegraphics[width=8.0cm,angle=0]{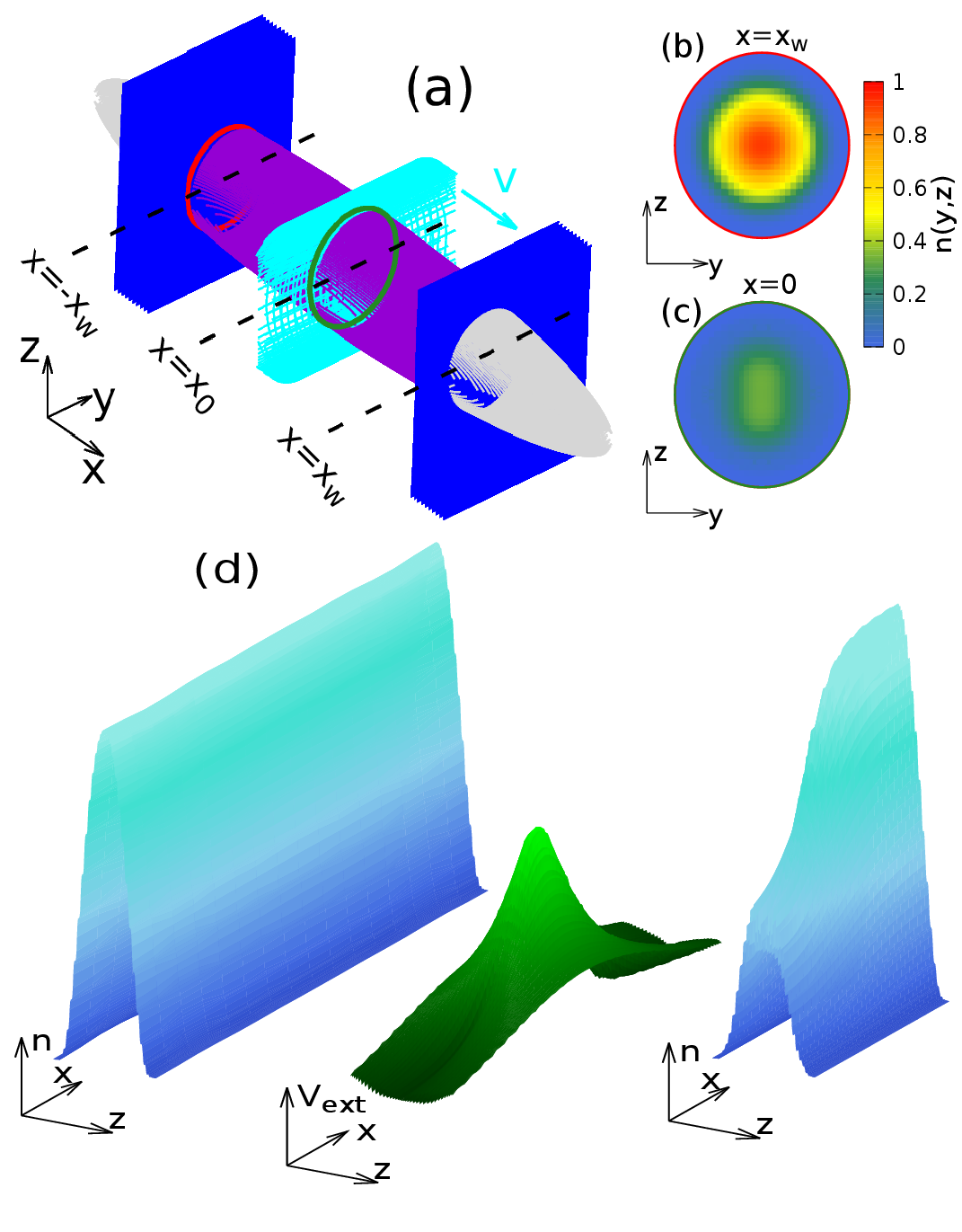}
\caption{(a) Atomic cloud (purple) with  two walls (dark blue) near its edges and a barrier (light blue) at its center.
              Contour maps of the number density, (b) at the position $x_{\mathrm{w}}$ of the wall (red circle) and (c) at the barrier center (green circle).
              (d) Left: Density profile in the absence of the barrier. Center: Typical shape of the barrier here considered. Right: Density profile in the presence of the barrier.}
\label{Figure-1}
\end{center} 
\end{figure}

In addition to a reliable account of the dynamics of pairing fluctuations in terms of the theoretical approach described above, what is needed for a correct interpretation of the experimental results of Refs.~\cite{Kwon-2020,DelPace-2021} is a detailed inclusion of the geometrical constraints there involved.
The experimental geometry utilized in these references is conveniently summarized in Fig.~\ref{Figure-1}, which provides details of the atomic cloud, the contour map of the number density, and
the density profiles both in the absence and in the presence of the Josephson barrier.
In the following, we shall refer to this figure when identifying the inhomogeneous environment in which the Fermi superfluid is embedded.

Specifically, the experimental geometry of Refs.~\cite{Kwon-2020,DelPace-2021} is reproduced schematically in Fig.~\ref{Figure-1}(a), where an atomic cloud initially with a strongly elongated ellipsoidal shape (purple) is affected by the raising of two walls (dark blue) near its edges (thus making the ellipsoid to almost resemble a cylinder) and of a barrier (light blue) at the center.
Both the trap and the barrier contribute to the external potential $V_{\mathrm{ext}}(\mathbf{r})$ that enters the local chemical potential $\mu(\mathbf{r}) = \mu - V_{\mathrm{ext}}(\mathbf{r})$.
In addition, Figs.~\ref{Figure-1}(b) and (c)  show the contour maps of the ensuing number density profiles, respectively, at the positions of the walls (red circle) and at the barrier center (green circle), thus showing how the presence of the barrier strongly reduces the density locally.
Finally, Fig.~\ref{Figure-1}(d)  reports, at the left the density profile integrated along the $y$ coordinate in the absence of the barrier, at the center the shape of the experimental barrier, and at the right the corresponding density profile in the presence of the barrier centered at $x=0$, shown for symmetry only along the positive side of the barrier (these plots cover only $1/7$ of the distance from the barrier center to the wall at positive $x$, where the density profiles recover their asymptotic values).
All these profiles are calculated by the mLPDA approach of Ref.~\cite{PPS-companion-PRB-article}. 

The trapping potential of Fig.~\ref{Figure-1}(a)  has the standard form of an anisotropic harmonic potential
\begin{equation}
V_{\mathrm{trap}}(x,y,z) = \frac{1}{2} m \left(\omega_{x}^{2} x^{2} + \omega_{y}^{2} y^{2} + \omega_{z}^{2} z^{2} \right) \, ,
\label{V-trap}
\end{equation}
with experimental values $\omega_{x,y,z} = 2 \pi(12,165,140)Hz$ in Ref.~\cite{Kwon-2020} and $\omega_{x,y,z} = 2 \pi (17,300,290)Hz$ in Ref.~\cite{DelPace-2021}. 
In both cases $\omega_{x} \ll \omega_{y} \simeq \omega_{z}$, such that the particle density acquires the cigar-shaped form shown in  Fig.~\ref{Figure-1}(a).

In addition, in the above experiments two walls where raised at positions $\pm x_{\mathrm{w}}$ along the major axis of the ellipsoid (as shown in   Fig.~\ref{Figure-1}(a)), 
where $k_{F}^{t} x_{\mathrm{w}} = 187.1$ \cite{Kwon-2020} and $k_{F}^{t} x_{\mathrm{w}} = 253.3$ \cite{DelPace-2021} in units of the trap Fermi wave vector $k_{F}^{t} = \sqrt{2m E_{F}^{t}}$ 
associated with the \emph{trap\/} Fermi energy $E_{F}^{t} = \omega_{0} (3N)^{1/3}$, where $\omega_{0}$ is the average trap frequency and $N$ the total number of fermionic ($^{6}\mathrm{Li}$) atoms
\emph{before\/} the raising of the walls (we set $\hbar = 1$ throughout).

As a consequence, in both cases the total external potential, acting on the fermionic atoms before the subsequent raising of the Josephson barrier, can be taken of the form:
\begin{eqnarray}
& V_{\mathrm{ext}}(x,y,z) & \, = V_{\mathrm{trap}}(x,y,z) 
\label{V-ext} \\
& + & \hspace{-0.6cm} 1.2 \times 10^{3} E_{F}^{t} \, \left\{ \theta[(k_{F}^{t}(x-x_{\mathrm{w}})] + \theta[-(k_{F}^{t}(x+x_{\mathrm{w}})] \right\} .
\nonumber
\end{eqnarray}
Here, the pre-factor multiplying the unit step functions is chosen large enough that the atoms cannot leak through the walls.
In this way, the total number of fermionic trapped atoms is reduced from $N = 2.6 \times 10^{5}$ \cite{Kwon-2020,footnote-Roati} and $N = 3.0 \times 10^{5}$ \cite{DelPace-2021,footnote-Roati} before the raising of the walls, 
to $N_{\mathrm{w}} = (1.0 \div 1.4) \times 10^{5}$ \cite{Kwon-2020,footnote-Roati} and $N_{\mathrm{w}} = 1.6 \times 10^{5}$ \cite{DelPace-2021,footnote-Roati} \emph{after\/} the raising of the walls.
These values of $N_{\mathrm{w}}$ are reported in Figs.~\ref{Figure-3} and \ref{Figure-4} below.

Finally, a Josephson barrier, raised at the center of the major axis of the ellipsoid (cf. Fig.~\ref{Figure-1}(d)), adds to $V_{\mathrm{ext}}(x,y,z)$ of Eq.~(\ref{V-ext}). 
This barrier is Gaussian along $x$, uniform along $y$, and decays with a linear power law for large $z$:
\begin{equation}
V_{\mathrm{barrier}}(x,y,z) = V_{0}(z) \, \exp\left(-2\dfrac{x^{2}}{w(z)^{2}}\right) 
\label{V-barrier} 
\end{equation}
with
\begin{equation}
V_{0}(z) = \frac{V_{0}}{\sqrt{1+\left(\frac{z}{z_{R}}\right)^{2}}}
\label{V_0}
\end{equation}
and
\begin{equation}
w(z) = w_{0} \, \sqrt{1+\left(\frac{z}{z_{R}}\right)^{2}} \, .
\label{w_0}
\end{equation}
In the experiments, the three parameters $(V_{0},z_{R},w_{0})$ take the following values: 
$V_{0}/E_{F}^{t} = (0.38,0.455,0.53)$ \cite{Kwon-2020} (corresponding to the panels of Fig.~\ref{Figure-3} below) and $V_{0}/E_{F}^{t} = 0.411$ \cite{DelPace-2021};
$k_{F}^{t} z_{R} = 14.24$ \cite{Kwon-2020} and $k_{F}^{t} z_{R} = 26.07$ \cite{DelPace-2021};
$k_{F}^{t} w_{0} = 2.54$ \cite{Kwon-2020} and $k_{F}^{t} w_{0} = 3.44$ \cite{DelPace-2021}.

\section{Josephson conditions} 
\label{sec:Josephson_conditions}

We now consider the specific treatment of the Josephson effect, whereby a steady current impinges on the fixed barrier of Fig.~\ref{Figure-1}(d), say, from negative $x$.

To mimic what occurs in the experimental setups of Refs.~\cite{Kwon-2020,DelPace-2021}, where the atomic cloud is at rest and the barrier steadily moves across it, we impose the condition that no current flows in the transverse ($y$ and $z$) directions. 
We have obtained this information from an independent numerical simulation performed in the BEC limit of the crossover with the time-dependent Gross-Pitaevskii equation \cite{Simonucci-unpublished}, 
whereby the current flow lines are seen not to bend away from the longitudinal ($x$) axis.
With this provision, the current flow can be treated as \emph{locally\/} uniform for given values of the transverse coordinates $y$ and $z$. 
This enables us to apply the methods developed in Ref.~\cite{Piselli-2020} to deal with the Josephson effect for a system which is fully homogeneous in the directions transverse to the current flow
(with the essential difference, however, that we now include explicitly pairing fluctuations over and above the approach of Ref.~\cite{Piselli-2020}).

In practice, owing to the negligible value of the density in the outer edge of the truncated ellipsoid of Fig.~\ref{Figure-1}(a), we further neglect the slight transverse bulge and assimilate 
the truncated ellipsoid to a cylinder.
This cylinder is then partitioned into a bundle of (at most 441) tubular filaments, in each one of which the approach of Ref.~\cite{Piselli-2020} is locally implemented, with boundary conditions specified by the local values of the gap parameter and density at the positions of the walls in Fig.~\ref{Figure-1}(a).
Finally, by fixing the difference $\delta \phi$ of the phase of the gap parameter between the walls in Fig.~\ref{Figure-1}(a), which is due to the presence of the barrier, the ``local''
Josephson characteristics $j$ vs $\delta \phi$ for the current density are calculated for each tubular filament and then integrated over all filaments across the transverse directions.
This procedure yields eventually the ``global'' Josephson characteristic $I(\delta \phi)$ for the total current $I$.

\begin{figure}[t]
\begin{center}
\includegraphics[width=8.5cm,angle=0]{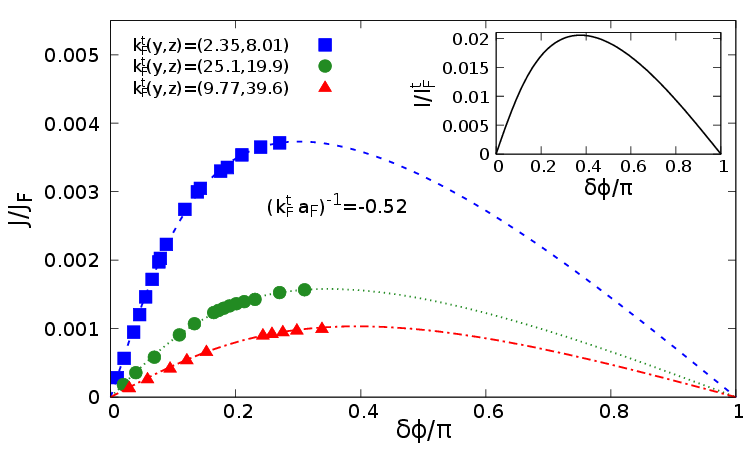}
\caption{The local Josephson characteristics for the current density are shown for three filaments with transverse coordinates $y$ and $z$, with the trap coupling parameter $(k_{F}^{t} a_{F})^{-1}=-0.52$ 
              and temperature $T/T_{F}^{t}=0.06$ where $T_{F}^{t}$ is the trap Fermi temperature corresponding to $E_{F}^{t}$.
              The inset shows the corresponding global Josephson characteristic for the total current.
              The local $j_{F} = k_{F}(x_{\mathrm{w}},y,z) n(x_{\mathrm{w}},y,z)/m$ normalizes the current density $j$ for each filament ($m$ being the fermion mass), while the global  
              $I_{F}^{t} = k_{F}^{t} \int \!\! dy dz \, n(x_{\mathrm{w}},y,z)/m$ normalizes the total current $I$ (needed when comparing with the experimental data).} 
\label{Figure-2}
\end{center} 
\end{figure} 

A typical example of the above procedure is shown in Fig.~\ref{Figure-2}  for the value $-0.52$ of trap coupling parameter $(k_{F}^{t} a_{F})^{-1}$ on the BCS side of the crossover and the
temperature $T/T_{F}^{t}=0.06$, with the barrier corresponding to Fig.~\ref{Figure-3} below.
Three different local Josephson characteristics are shown, corresponding to the filaments specified by the transverse coordinates $y$ and $z$ reported in the figure,
which have been selected in order to differentiate the corresponding Josephson characteristics as much as possible.
In addition, the dashed lines correspond to the universal fitting function for the Josephson characteristics given by Eq.~(14) of Ref.~\cite{Piselli-2020}, which is of help in drawing the ``right” branches of the Josephson characteristics 
(known to be unstable - cf. Sec.~V of Ref.~\cite{Spuntarelli-2010}). 
Recall that, not only the heights and widths of the barrier, but also the local Fermi wave vectors associated with the local density $n(x_{\mathrm{w}},y,z)$ are different for each filament (where $x_{\mathrm{w}}$ is the
coordinate specified in Fig.~\ref{Figure-1}(a)  above).
Changes of these quantities from an inner to an outer filament are expected to have different effects on the shape of the local Josephson characteristics.

As it was shown in Fig.~5 of Ref.~\cite{Piselli-2020}, where the effects on the Josephson characteristics due to changes of the barrier height and width were disentangled from each other,
an increase of either the barrier height or width shifts the maximum of the Josephson characteristics to the right, while changes of the Fermi wave vector do not provide clear indications to where this maximum would shift. 
This is even more so in the present trapped case, where the values of the local Fermi wave vector and the barrier height are maximum at the trap center and decrease away from it, 
while the width of the barrier is minimum at the trap center and increases away from it.
To the extent that it is not possible to disentangle these effects from each other, only the overall shape of the global Josephson characteristic for the total current is physically meaningful.
This global Josephson characteristic is what is shown in the inset of Fig.~\ref{Figure-2}. 

\section{Comparison with experiments} 
\label{sec:comparison_experim} 

Quite generally, an important piece of information that can be extracted from the Josephson characteristics is the value of the critical current $I_{c}$, 
which corresponds to the maximum value of the supercurrent that is able to flow  across a  given barrier.
The experiments reported in Refs.~\cite{Kwon-2020,DelPace-2021} have obtained values of $I_{c}$, respectively, at low temperature for several barriers and various couplings across the BCS-BEC crossover
and as a function of temperature at unitarity.
We are now in a position to compare in detail the experimental values for the critical current given in Refs.~\cite{Kwon-2020,DelPace-2021} with our numerical results for $I_{c}$,
obtained by calculating the Josephson characteristics for the total current which adds the contributions from the local currents carried by the tubular filaments as described above.
This comparison will enable us to obtain  a stringent test on our  theoretical approach.

\begin{figure}[t]
\begin{center}
\includegraphics[width=6.8cm,angle=0]{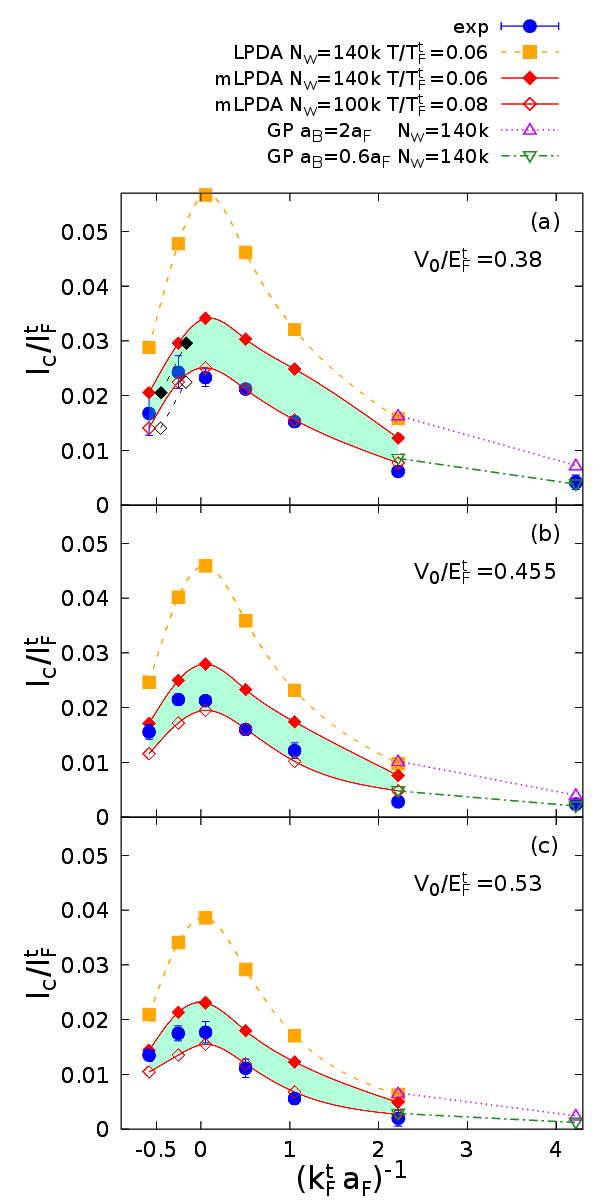}
\caption{Critical current $I_{c}$ (in units of of $I_{F}^{t}$) vs the trap coupling $(k_{F}^{t} a_{F})^{-1}$, for three barriers with the  same width and different heights.
              The experimental data from Ref.~\cite{Kwon-2020} (dots with error bars) are compared with the theoretical results obtained by solving both LPDA and mLPDA equations.
              The theoretical results are obtained for different temperatures and different values of the total number of atoms $N_{\mathrm{w}}$ contained in the atomic cloud
              after raising the walls, which correspond to the experimental ranges of temperature and $N_{\mathrm{w}}$ \cite{Kwon-2020,footnote-Roati}.
              Consideration of these ranges gives rise to the shaded areas spanned by the numerical calculations.
              On the BEC side, where the GP equation applies \cite{PS-2003}, results obtained with bosonic scattering lengths $a_{B} = 2.0 a_{F}$ (magenta triangles) and $a_{B} = 0.6 a_{F}$   
              (green triangles) are reported.
              In panel (a), the filled and empty black diamonds correspond to a simplified version of the extended GMB approach (see the text).}             
\label{Figure-3}
\end{center} 
\end{figure} 

Figure~\ref{Figure-3} shows an extensive comparison between the coupling dependence  of the critical current $I_{c}$ obtained experimentally in Ref.~\cite{Kwon-2020} and theoretically  by solving  the LPDA and mLPDA equations, that is, without and with the inclusion of beyond-mean-field pairing fluctuations, respectively \cite{footnote-errata-corrige}.
Here, we have followed the convention of Ref.~\cite{Kwon-2020} and identified the coupling in terms of the Fermi wave vector $k_{F}^{t} = \sqrt{2m E_{F}^{t}}$ associated with the \emph{trap\/} Fermi energy $E_{F}^{t} = \omega_{0} (3N)^{1/3}$, where $N$ is the total number of atoms and $\omega_{0}$ the average trap frequency associated with the ellipsoidal-shaped atomic cloud before the walls are raised 
(cf. Fig.~\ref{Figure-1} above).
Correspondingly, $I_{F}^{t}$ is the global current defined in terms of $k_{F}^{t}$ and the total particle density integrated over the spatial directions transverse to the current flow 
(cf. Fig.~\ref{Figure-2} above)).
This comparison is reported  at low temperature for three barriers with increasing height and for several couplings from the BCS to the BEC side of unitarity.

The overall agreement,  obtained over a quite extended range of coupling, between the results of the mLPDA calculations and the experimental data appears rather remarkable.
It points out, in particular, the crucial role played by pairing fluctuations in the crossover region, thereby explicitly validating the mLPDA approach over wide physical conditions.
Figure~\ref{Figure-3}  also shows the results obtained by an independent calculation with the GP equation for composite bosons in the BEC limit of the crossover, 
to which both the LPDA and mLPDA equations reduce  in this limit,
by considering either the exact value $a_{B} = 0.6 a_{F}$ of the bosonic scattering length \cite{Brodsky-2006} or its Born approximation value $a_{B} = 2.0 a_{F}$ \cite{PS-2003}.

 A distinctive feature of many-body diagrammatic approaches (like the one we adopt here) is that, being modular in nature, they are amenable to improvement by adding diagrammatic terms 
relevant to the physics of the problem at hand (provided, of course, they can be implemented with reasonable numerical efforts).
In the present context, these diagrams are related to  the extended GMB approach of Ref.~\cite{Pisani-2018}, whose importance has recently been certified in different experimental contexts, both at low temperature in the superfluid phase \cite{Moritz-2022} and at the critical temperature \cite{Koehl-2023}.
Full implementation of the extended GMB approach, however, would not only require us to include in the LPDA equation (\ref{LPDA-differential-equation}) (a local version of) 
the bosonic-like self-energy terms identified in Ref.~\cite{Pisani-2018}, but also to calculate them in the presence of a supercurrent.
This program exceeds the objectives of the present work. 
Nevertheless, we give here a proof-of-principle for the role played by the extended GMB approach in the present context, by adopting the simplified procedure described in Sec.~IV-D  of Ref.~\cite{PPS-companion-PRB-article}.
The ensuing numerical results are shown in Fig.~\ref{Figure-3}(a) by the position of the black filled and empty diamonds, 
which delimit the experimental data better than the original red filled and empty diamonds obtained by the mLPDA approach,
thus improving the comparison with the experiment. 

\begin{figure}[t]
\begin{center}
\includegraphics[width=8.9cm,angle=0]{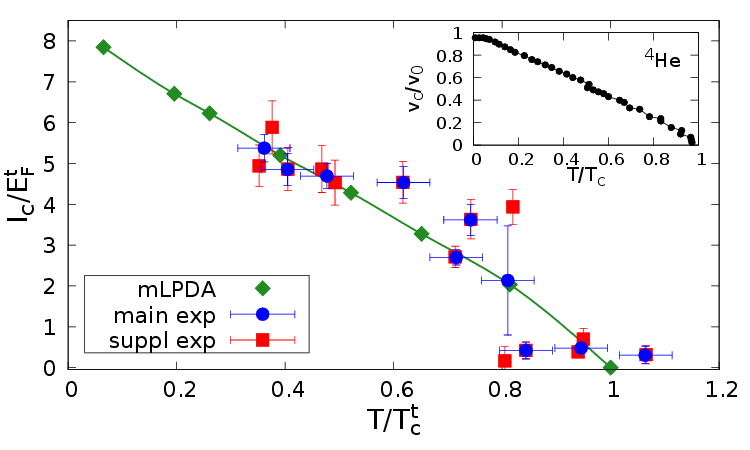}
\caption{Temperature dependence (in units of the trap critical temperature $T_{c}^{t}$) of the critical current $I_{c}$ (in units of $E_{F}^{t}$) at unitarity.
              The theoretical results obtained by the mLPDA approach with $N_{\mathrm{w}} = 160k$  (green diamonds) are compared with the experimental data,
              taken both  from the main text (blue dots with error bars) and from the supplemental material (red squares with error bars) of Ref.~\cite{DelPace-2021}.
              The inset shows the critical velocity of superfluid $^{4}\mathrm{He}$ (normalized to its zero-temperature value), as extracted  from Fig.~12 of Ref.~\cite{Varoquaux-2015}.}
\label{Figure-4}
\end{center} 
\end{figure} 

Finally, Fig.~\ref{Figure-4} compares the temperature dependence of the critical current $I_{c}$ obtained experimentally at unitarity in Ref.~\cite{DelPace-2021} with the theoretical results of the mLPDA equation.
This comparison is shown over the full temperature range, from low temperature up to the trap critical temperature $T_{c}^{t}$. 
Here, for internal consistency the theoretical value of $T_{c}^{t}$ is calculated like in Ref.~\cite{PPPS-2004}, while, lacking a reliable experimental estimate for $T_{c}^{t}$, the temperatures of the experimental data 
are normalized to the value of $T_{c}^{t}$ obtained by the fully-self-consistent $t$-matrix calculation of Ref.~\cite{Pini-2020}
Even in this case, the agreement between the experimental data and the results of the mLPDA approach  appears extremely good.
A feature to  be emphasized is the \emph{linear\/} trend of the theoretical results, which appears to be consistent with the experimental data.
In Ref.~\cite{PPS-companion-PRB-article} this linear behavior is shown to be intermediate between  a (slightly) convex behavior in the BCS regime and  a (slightly) concave behavior in the BEC regime.
In addition, the inset of Fig.~\ref{Figure-4}  shows that a linear temperature behavior is shared by the critical velocity of superfluid $^{4}\mathrm{He}$, as taken from Fig.~12 of Ref.~\cite{Varoquaux-2015}.
The similarity between the behavior of the unitary Fermi gas and $^{4}\mathrm{He}$ has been repeatedly noted over the years, ranging from the superfluid fraction \cite{Sidorenkov-2013}
to  the  sound propagation \cite{Kuhn-2020}.

\section{Conclusions} 
\label{sec:conclusions} 
 
In this article, we have considered the  so far   unsettled issue of combining the many-body dynamics of beyond-mean-field pairing fluctuations, which is relevant to 
a fermionic superfluid undergoing the BCS-BEC crossover, with the presence of nontrivial geometrical constraints that may substantially affect the superfluid flow.
In the case of the Josephson effect, we have succeeded in dealing  with these two aspects \emph{on the same footing\/}, by implementing the novel mLPDA  approach developed in Ref.~\cite{PPS-companion-PRB-article}.
The favorable comparison of our numerical results, with the recently available experimental data in ultra-cold Fermi gases  under a variety of circumstances, 
should accordingly be regarded as a stringent test for the validity of  the approach of Ref.~\cite{PPS-companion-PRB-article}.
In addition, this approach, being based on the many-body Green’s functions theory, offers further  perspectives for improvement, by adding diagrammatic contributions over and above
those explicitly considered in Ref.~\cite{PPS-companion-PRB-article}.
A preliminary test along these lines has already provided promising results.


\begin{center}
\begin{small}
{\bf ACKNOWLEDGMENTS}
\end{small}
\end{center}

 We are indebted to G. Roati for support and discussions, and to G. Del Pace for a critical reading of the manuscript.
 
\newpage
\appendix   
\vspace{1.5cm}
\section{A SHORT SUMMARY ABOUT \\ THE LPDA and mLPDA APPROACHES}
\label{sec:Appendix-A}

The LPDA differential equation for the (complex) gap parameter $\Delta(\mathbf{r})$ was introduced in Ref.~\cite{Simonucci-2014}. It reads:
\begin{equation}
\left[ \frac{m}{4 \pi a_{F}} + \mathcal{I}_{0}(\mathbf{r}) +  
\mathcal{I}_{1}(\mathbf{r}) \! \left( \frac{\nabla^{2}}{4m} - i \, \frac{\mathbf{A}(\mathbf{r})}{m} \cdot \nabla \! \right) \right]\! \Delta(\mathbf{r}) = 0                                                
\label{LPDA-differential-equation-bis}
\end{equation}
\noindent
where $m$ is the fermion mass, $a_{F}$ the fermionic scattering length, $\mathbf{A}(\mathbf{r})$ the vector potential, and
\begin{equation}
\mathcal{I}_{0}(\mathbf{r}) = \int \! \frac{d \mathbf{k}}{(2 \pi)^{3}} \, 
\left\{ \frac{ 1 - 2 f_{F}(E_{+}^{\mathbf{A}}(\mathbf{k}|\mathbf{r})) }{2 \, E(\mathbf{k}|\mathbf{r})} - \frac{m}{\mathbf{k}^{2}} \right\}
\label{I_0-definition-finite_temperature}
\end{equation}
\noindent
and
\begin{eqnarray}
\mathcal{I}_{1}(\mathbf{r}) & = & \frac{1}{2} \, \int \! \frac{d \mathbf{k}}{(2 \pi)^{3}} 
\left\{  \frac{\xi(\mathbf{k}|\mathbf{r})}{2 \, E(\mathbf{k}|\mathbf{r})^{3}} \, \left[  1 - 2 f_{F}(E_{+}^{\mathbf{A}}(\mathbf{k}|\mathbf{r})) \right] \right.                          
\nonumber \\
& + &  \frac{\xi(\mathbf{k}|\mathbf{r})}{E(\mathbf{k}|\mathbf{r})^{2}} \, 
\frac{\partial f_{F}(E_{+}^{\mathbf{A}}(\mathbf{k}|\mathbf{r}))}{\partial E_{+}^{\mathbf{A}}(\mathbf{k}|\mathbf{r})} 
\label{I_1-definition-finite_temperature} \\
& - & \left. \frac{\mathbf{k}\cdot\mathbf{A}(\mathbf{r})}{\mathbf{A}(\mathbf{r})^{2}} \, \frac{1}{E(\mathbf{k}|\mathbf{r})} \, 
\frac{\partial f_{F}(E_{+}^{\mathbf{A}}(\mathbf{k}|\mathbf{r}))}{\partial E_{+}^{\mathbf{A}}(\mathbf{k}|\mathbf{r})} \right\} \, .
\nonumber
\end{eqnarray}
\noindent
In these expressions,  
$\xi(\mathbf{k}|\mathbf{r}) = \frac{\mathbf{k}^{2}}{2m} - \bar{\mu}(\mathbf{r})$,
$E(\mathbf{k}|\mathbf{r}) = \sqrt{\xi(\mathbf{k}|\mathbf{r})^{2} + |\Delta(\mathbf{r})|^{2}}$, and
$E_{+}^{\mathbf{A}}(\mathbf{k}|\mathbf{r}) = E(\mathbf{k}|\mathbf{r}) - \frac{\mathbf{k} \cdot \mathbf{A}(\mathbf{r})}{m}$,
where the local chemical potential $\bar{\mu}(\mathbf{r}) = \mu - V_{\mathrm{ext}}(\mathbf{r}) - \mathbf{A}(\mathbf{r})^{2}/(2m)$ accounts for the presence of an external potential $V_{\mathrm{ext}}(\mathbf{r})$.

The LPDA equation (\ref{LPDA-differential-equation}) has to be supplied by the expressions for the local number density $n(\mathbf{r})$ and current density $\mathbf{j}(\mathbf{r})$.
They read \cite{Simonucci-2014}:
\begin{equation}
n(\mathbf{r}) =  \int \! \frac{d\mathbf{k}}{(2 \pi)^{3}} \left\{ 1 - 
\frac{\xi^{\mathbf{A}}(\mathbf{k}|\mathbf{r})}{E^{\mathbf{A}}(\mathbf{k}|\mathbf{r})} 
\left[ 1 - 2 f_{F}(E^{\mathbf{A}}_{+}(\mathbf{k}|\mathbf{r})) \right] \right\} 
\label{number-density-mf}
\end{equation}
\noindent
and
\begin{eqnarray}
\mathbf{j}(\mathbf{r}) & = & \frac{1}{m} \! \left(\nabla \phi(\mathbf{r}) - \mathbf{A}(\mathbf{r}) \right) \, n(\mathbf{r})
\nonumber \\
& + &  2 \int \! \frac{d\mathbf{k}}{(2 \pi)^{3}} \, \frac{\mathbf{k}}{m} \, f_{E} \! \left( E^{\mathbf{A}}_{+}(\mathbf{k}|\mathbf{r}) \right) \, .
\label{current-density-mf} 
\end{eqnarray}
In these expressions,
\begin{eqnarray}
\xi^{\mathbf{A}}(\mathbf{k}|\mathbf{r}) & = & \frac{\mathbf{k}^{2}}{2m} - \mu(\mathbf{r}) + 
\frac{1}{2m} \left( \nabla \phi(\mathbf{r}) - \mathbf{A}(\mathbf{r}) \right)^{2} \, ,
\nonumber \\
E^{\mathbf{A}}(\mathbf{k}|\mathbf{r}) & = & \sqrt{\xi^{\mathbf{A}}(\mathbf{k}|\mathbf{r})^{2}
+ |\Delta(\mathbf{r})|^{2}}  \, ,
\nonumber \\
E^{\mathbf{A}}_{+}(\mathbf{k}|\mathbf{r}) & = & E^{\mathbf{A}}(\mathbf{k}|\mathbf{r}) 
+ \frac{\mathbf{k}}{m} \cdot \left( \nabla \phi(\mathbf{r}) - \mathbf{A}(\mathbf{r}) \right) 
\label{three-definitions}
\end{eqnarray}
where now $\mu(\mathbf{r}) = \mu - V_{\mathrm{ext}}(\mathbf{r})$ contains only the external potential and $2 \phi(\mathbf{r})$ is the phase of the gap parameter such that 
$\Delta(\mathbf{r}) = |\Delta(\mathbf{r})| e^{i 2 \phi(\mathbf{r})}$.

For the Josephson effect of concern in the present article, $\mathbf{A}(\mathbf{r}) \rightarrow - \mathbf{Q}_{0}$ where $\mathbf{Q}_{0}$ accounts the superfluid flow before a barrier is raised 
to split the fermionic superfluid in two (left and right) parts \cite{Piselli-2020}.
In this case, $\phi(\mathbf{r}) \rightarrow \mathbf{Q}_{0} \cdot \mathbf{r} + \phi(\mathbf{r})$ in the phase of the gap parameter, where $\phi(\mathbf{r})$ at the right side is now the \emph{additional\/} phase
contributed by the presence of the barrier.

The expressions (\ref{number-density-mf}) and (\ref{current-density-mf}) hold within a mean-field decoupling.
 They can, however, be modified to include pairing fluctuations (pf) beyond mean field.
This can conveniently be done by introducing a \emph{local version\/} $\mathcal{G}_{11}^{\mathrm{pf}}(\mathbf{k},\omega_{n};\mathbf{q}|\mathbf{r})$ of the diagonal (normal) single-particle Green's function 
$\mathcal{G}_{11}^{\mathrm{pf}}(\mathbf{k},\omega_{n};\mathbf{q})$, which has, in turn, to account in a consistent way for the presence of a superfluid flow specified by the wave vector $\mathbf{q}$.
This Green's function reads \cite{PPS-companion-PRB-article}:
\begin{widetext}
\begin{large}
\begin{equation}
\mathcal{G}_{11}^{\mathrm{pf}}(\mathbf{k},\omega_{n};\mathbf{q}) = 
\frac{1}{ i \omega_{n} - \xi(\mathbf{k}+\mathbf{q}) - \mathfrak{S}_{11}^{\mathrm{pf}}(\mathbf{k},\omega_{n};\mathbf{q}) - \frac{\Delta_{\mathbf{q}}^{2}}{i \omega_{n} + \xi(\mathbf{k}-\mathbf{q}) 
+ \mathfrak{S}_{11}^{\mathrm{pf}}(\mathbf{k},-\omega_{n};\mathbf{q})}} 
\label{G-11-pairing-flucuations_and_current}
\end{equation}
\end{large}
with the diagonal (normal) single-particle self-energy 
\begin{equation}
\mathfrak{S}_{11}^{\mathrm{pf}}(\mathbf{k},\omega_{n};\mathbf{q}) \! = \! - \! \int \!\!\! \frac{d \mathbf{Q}}{(2 \pi)^{3}} \, \frac{1}{\beta} \sum_{\nu} 
                                                              \Gamma_{11}(\mathbf{Q},\Omega_{\nu};\mathbf{q}) \, \mathcal{G}_{11}^{\mathrm{mf}}(\mathbf{Q} - \mathbf{k},\Omega_{\nu}-\omega_{n};\mathbf{q}) \, .
\label{single-particle self-energy}
\end{equation}
\end{widetext}
In these expressions, $\xi(\mathbf{k}) = \mathbf{k}^{2}/(2m) - \mu$, $\beta = (k_{B} T)^{-1}$ is the inverse temperature ($k_{B}$ being the Boltzmann constant), $\omega_{n} = (2n+1)\pi/\beta$ ($n$ integer)
and $\Omega_{\nu} =  2\nu\pi/\beta$ ($\nu$ integer) are fermionic and bosonic Matsubara frequencies \cite{FW-1971}, respectively, and $\Delta_{\mathbf{q}}$ is the associated magnitude of the gap parameter.
In addition, $\Gamma_{11}(\mathbf{Q},\Omega_{\nu};\mathbf{q})$ is the diagonal element of the particle-particle ladder in the broken-symmetry phase, which has also to take into account the presence of a 
superfluid flow \cite{PPS-companion-PRB-article}.

With a suitable prescription to obtain from the form (\ref{G-11-pairing-flucuations_and_current}) of $\mathcal{G}_{11}^{\mathrm{pf}}(\mathbf{k},\omega_{n};\mathbf{q})$ its local version 
$\mathcal{G}_{11}^{\mathrm{pf}}(\mathbf{k},\omega_{n};\mathbf{q}|\mathbf{r})$ (for whose details we refer to the companion article \cite{PPS-companion-PRB-article}),
the expressions for the local particle density and current that include pairing fluctuations beyond mean field read:
\begin{eqnarray}
n(\mathbf{r}) & = & \frac{2}{\beta} \sum_{n} e^{i \omega_{n} \eta} \! \int \!\!\! \frac{d \mathbf{k}}{(2 \pi)^{3}} \,\, \mathcal{G}_{11}^{\mathrm{pf}}(\mathbf{k},\omega_{n};\mathbf{q}|\mathbf{r})
\label{local-density-pairing-fluctuations} \\
\mathbf{j}(\mathbf{r}) & = & \frac{1}{m} \left( \mathbf{Q_{0}} + \nabla \phi(\mathbf{r}) \right) n(\mathbf{r}) 
\nonumber \\
& + & \frac{2}{\beta} \sum_{n} e^{i \omega_{n} \eta} \! \int \!\!\! \frac{d \mathbf{k}}{(2 \pi)^{3}} \frac{\mathbf{k}}{m} \, \mathcal{G}_{11}^{\mathrm{pf}}(\mathbf{k},\omega_{n};\mathbf{q}|\mathbf{r})
\label{local-current-pairing-fluctuations}
\end{eqnarray}
$\eta$ being a positive infinitesimal.
These expressions reduce to the LPDA results (\ref{number-density-mf}) and (\ref{current-density-mf}) when the self-energy (\ref{single-particle self-energy}) is set to zero in the expression (\ref{G-11-pairing-flucuations_and_current})
of the Green's function.

The (modified) mLPDA approach, introduced in Ref.~\cite{PPS-companion-PRB-article} and utilized in the present article to account for the experimental results of Refs.~\cite{Kwon-2020,DelPace-2021}, is obtained 
by supplementing the LPDA differential equation (\ref{LPDA-differential-equation}) with the expressions (\ref{local-density-pairing-fluctuations}) and (\ref{local-current-pairing-fluctuations}) for the local particle density and current, in the place of the expressions (\ref{number-density-mf}) and (\ref{current-density-mf}) that were originally used in the LPDA approach in Ref.~\cite{Simonucci-2014}.

	


\begin{thebibliography}{99}
\bibitem{footnote-phase} The only place where $\nabla \phi(\mathbf{r})$ does not appear is in the local expression of the particle-particle ladder (cf. Ref.\cite{PPS-companion-PRB-article}).
\end{thebibliography}
\end{document}